\documentclass{article}

\pdfoutput=1

\usepackage[preprint, nonatbib]{neurips_2020}

\usepackage[sortcites=False, backend=biber, autocite=superscript, maxbibnames=99, doi=false, arxiv=false, url=false, sorting=none, doi=false,isbn=false, style=numeric-comp]{biblatex}

\DefineBibliographyStrings{english}{%
  backrefpage = {page},%
  backrefpages = {pages},%
}
\bibliography{neurips_2020.bib}

\usepackage[utf8]{inputenc} 
\usepackage[T1]{fontenc}    
\usepackage{hyperref}       
\usepackage{url}            
\usepackage{booktabs}       
\usepackage{amsfonts}       
\usepackage{nicefrac}       
\usepackage{microtype}      
\usepackage{makecell}

\usepackage{amsmath}
\usepackage{algorithm}
\usepackage{algorithmic}
\usepackage{amsmath}
\usepackage{caption}
\usepackage{graphicx}
\usepackage[center]{subfigure} 
\usepackage{amssymb}
\usepackage{color}
\usepackage{threeparttable}
\usepackage{multirow}
\usepackage{multicol}
\usepackage[table,xcdraw]{xcolor}
\usepackage{colortbl}
\usepackage{wrapfig}  

\usepackage[title]{appendix}

\usepackage{authblk}

\title{AUSN: Approximately Uniform Quantization by Adaptively Superimposing Non-uniform Distribution for Deep Neural Networks}

\author[1]{Fangxin Liu}
\author[2]{Wenbo Zhao}
\author[3]{Yanzhi Wang}
\author[4]{Changzhi Dai}
\author[1,5]{Li Jiang \thanks {Corresponding author: jiangli@cs.sjtu.edu.cn}\,\ }
\affil[1]{School of Electronic Information and Electrical Engineering, Shanghai Jiao Tong University}
\affil[2]{SJTU-UM Joint Institute, Shanghai Jiao Tong University}
\affil[3]{Northeastern University}\affil[4]{DeepBlue Technology (Shanghai) Co., Ltd.}
\affil[5]{MoE Key Lab of Artificial Intelligence, AI Institute, Shanghai Jiao Tong University}

\begin{document}
	
	\maketitle
	
	\begin{abstract}
		Quantization is essential to simplify DNN inference in edge applications. Existing uniform and non-uniform quantization methods, however, exhibit an inherent conflict between the \emph{representing range} and \emph{representing resolution}, and thereby result in either underutilized bit-width or significant accuracy drop. Moreover, these methods encounter three drawbacks: i) the absence of a quantitative metric for in-depth analysis of the source of the quantization errors; ii) the limited focus on the image classification tasks based on CNNs; iii) the unawareness of the real hardware and energy consumption reduced by lowering the bit-width. 
		In this paper, we first define two quantitative metrics, i.e., the \emph{Clipping Error} and \emph{rounding error}, to analyze the quantization error distribution. We observe that the boundary- and rounding- errors vary significantly across layers, models and tasks. Consequently, we propose a novel quantization method to quantize the weight and activation. The key idea is to Approximate the Uniform quantization by \emph{Adaptively} \emph{Superposing} multiple Non-uniform quantized values, namely AUSN. AUSN is consist of a decoder-free coding scheme that efficiently exploits the bit-width to its extreme, a superposition quantization algorithm that can adapt the coding scheme to different DNN layers, models and tasks without extra hardware design effort, and a rounding scheme that can eliminate the well-known bit-width overflow and re-quantization issues. 
		Theoretical analysis~(see Appendix A) and accuracy evaluation on various DNN models of different tasks show the effectiveness and generalization of AUSN. The synthesis~(see Appendix B) results on FPGA show $2\times$ reduction of the energy consumption, and $2\times$ to $4\times$ reduction of the hardware resource.
	\end{abstract}
	
	\section{Introduction}
	Deploying DNN on edge devices, such as Internet-of-Things devices or mobile phones, is challenging on account of the limited computing and storage resources and energy budge provided by these edge devices. For instance, it takes 16 seconds on a mobile to complete an image recognition using VGG16~\cite{simonyan2014very}, which is intolerable for most applications~\cite{mao2017modnn}. Therefore, it is essential to compress the DNN for lower storage requirements and simpler arithmetic operations. 
	
	Among various compression techniques, quantization maps the weight values distributed in the infinite space of real numbers to the finite space of discrete numbers. Existing quantization methods, however, mainly encounter several issues. Uniform quantization, such as INT8~\cite{jacob2018quantization}, uses an affine function to map real value weights to uniformly distributed integers. This method results in a constant distance between the two adjacent quantized numbers, which denoted as~\emph{Representing Resolution}. The representing resolution becomes coarse as the bit-width decreases, which degrades the model accuracy. Non-uniform quantization, such as power-of-two~\cite{zhou2017incremental}, maps the weight values to exponential space. Non-uniform quantization has a extremely large \emph{Representing Range} of real number with low bit-width exponents. However, both of them can not balance the relationship between \emph{Representing Resolution} and \emph{Representing Range} using a low bit-width. For example, the \emph{Representing Resolution} of non-uniform distribution is too coarse when the exponent number is large and results in a significant quantization error. 
	
	
	Then, we find that some low bit-width quantization methods have poor generality. They cannot apply to the tasks like object detection and sequential network. For example, LSQ~\cite{esser2019learned} only quantizes the weights representing important image features in the convolution layer. Thus, this method only works for CNN based classification tasks. 
	
	Most quantization methods presume that lower bit-width can achieve higher speed up or smaller hardware consumption. The introduced decoder, however, may cause significant hardware and energy overhead. Various accelerator architectures may need different quantization bit-width. For instance, Digital-Signal-Processor~(DSP) and CPU with AVX prefer 4-bit and 8-bit quantization because the fundamental Multiply-and-accumulate~(MAC) can process 4-bit data. It is desired to build a relationship between the quantization bit-width and the real hardware/energy consumption. Such relationship can guide an efficient quantization given a specific accelerator architecture.

	In this work, First, AUSN, what we prose is not simply an algorithm, but a framework, including algorithm, coding, rounding, adaptively adjusting scheme. Second, we follow the principle of minimum accuracy drop and extremely small bit-width, to further salvage redundant bit-width to gain more quantization choices. Contributions of this paper are as follows:

	\begin{itemize}
		\item [1)] 
		\vspace{-0.5em}
		We first combines these two quantization methods and proposed a coding scheme without the decoder. The bit-width is divided into two parts, which are the symbol and data. The data part can be further divided into ``basic part'', and ``subdivision part'' for the superposition, which determine the~\emph{Representing Range} and ~\emph{Representing Resolution}, respectively. Then, we use the superposition of multiple numbers non-uniformly quantized with extremely small bit-width, to represent the weights and activations.
		\item [2)]
		We explore in-depth reasons for the quantization error in existing quantization methods: Clipping Error and rounding errors. According to these errors, we propose an adaptively adjusting scheme to reallocate the bit-width according to the weight distribution, reducing the quantization error.
		\item [3)]
		We design the rounding scheme to solve the problems about the extra computation effort and hardware overhead caused by overflow of bit-width and re-quantization.
		\item [4)]
    	We verify the quantized CNNs, sequential networks, and Yolov3-tiny by AUSN on various datasets without retraining and achieve excellent results that exceed state-of-the-art accuracy, and we prove the savings brought by AUSN in hardware resources and energy consumption on FPGA (see Appendix B). 
    	\item [5)]
    	We measure the effect of quantization on the model through information loss from the information theory. We also disscuss the tradeoff between quantization methods and performance of hardware, which can guide the design of accelerator (see Appendix A).
	\end{itemize}
	
	\section{Background}
	\subsection{Uniform quantization}
	
	Uniform quantization defines a \emph{Representing Range} $[\min, \max]$ and quantize weights in the range by first applying an affine function on them and then rounding to the closest integer.
	For example, all the floating-point weights are quantized to an integer in $[0, 255]$ in INT8 quantization~\cite{jacob2018quantization}. 
	
	If the original weights of the DNN exceed the \emph{Representing Range}, they will be quantized to the boundary values; if these quantized weights are of significant importance, the DNN model may suffer from substantial accuracy loss. {Uniform quantization ~\cite{jacob2018quantization,migacz20178} is widely used both in academia and industry because of the guarantee of high accuracy. For example, many open-source frameworks~(e.g., Pytorch, TensorFlow, etc.) support INT8 quantization. Although the \emph{Representing Range} of INT8 is broad enough for most DNN models, uniform quantization suffers drastic accuracy drop for DNN model with lower bit-width. The fewer quantized numbers are available to cover the whole weights, the farther apart from each other, consequently, which leads to a significant quantization error.}
	
	
	
	\subsection{Non-uniform quantization}
	The non-uniform quantization methods~\cite{ye2018unified,mcdanel2019full}, similar to clustering, aims at finding some ``centers" (i.e., the average value) that	can represent the majority of the weight values in the original weight distribution. Therefore, \textcolor{black}{these quantzaition methods need to store these ``centers", and the quantized number stores the index of ``centers" (such as Deep Compression~\cite{han2016deep}), which causes an obvious resource cost.}
	
	\textcolor{black}{Power-of-two quantization~\cite{johnson2018rethinking,mcdanel2019full} quantizes weights to the form of exponents (of power-of-two).} The quantized number symbolizes the exponent, $i.e.$, a 3-bit quantized value ``111'' means that the exponent is 7 and the original value is $2 ^ 7 = 128$. It has good \emph{Representing Range} since the range of quantized weights grows exponentially as the quantization bit-width increases. Besides, the power-of-two scheme is hardware-friendly: the multiplication operation can convert to a shift operation. For example, $\times 4$ can be substituted by shifting the bit sequence two bits to the left. \textcolor{black}{Such conversion is of great significance for the DNN hardware accelerator implemented in ASIC or FPGA~\cite{ding2019req}, because a tremendous amount of power- and resource-consuming DSPs in the multiplier can be substituted by the simple-and-effective Look-Up Tables~(LUTs). Take Xilinx Artix-7 as an example, this FPGA contains $53,200$ LUTs and $220$ DSP Slices, and the LUT resources are hundreds of times of the DSP, LUTs are a promising alternative to the implementation of multiplication.\footnote {https://china.xilinx.com/support/documentation/data\_sheets/ds190-Zynq-7000-Overview.pdf}. Such conversion can typically bring up to 50\% resource reductions on Xilinx FPGAs~\cite{faraone2019addnet}.}

	\section{Key idea of AUSN quantization}
	\textcolor{black}{How close the weights in a quantized model can approach its original value inherently determines its quantization error. We define two metrics to evaluate the quantization error. }
	\begin{itemize}
	\setlength{\itemsep}{0pt}
	\setlength{\parsep}{0pt}
    \setlength{\parskip}{0pt}
	    \item \textbf{\emph{Rounding Error}}: The error caused by rounding a weight to the nearest quantized value. 
	    \item \textcolor{black}{\textbf{\emph{Clipping Error}}: The error caused by clipping weight out of \emph{Representing Range} to extremum value.}
	\end{itemize}
	\begin{figure}[t!] 
		\centering  
		\setlength{\belowcaptionskip}{-0.5em}  
		\setlength{\abovecaptionskip}{-0.1em}  
		\subfigure[{\small An example showing the shortcoming of existing quantization}]{
			\label{fig:po2-shortcoming}
			\begin{minipage}[t]{0.45\linewidth}
				\centering
				\fbox{\rule[-.5pt]{0pt}{3.8cm} 
					\includegraphics[width=0.9\linewidth]{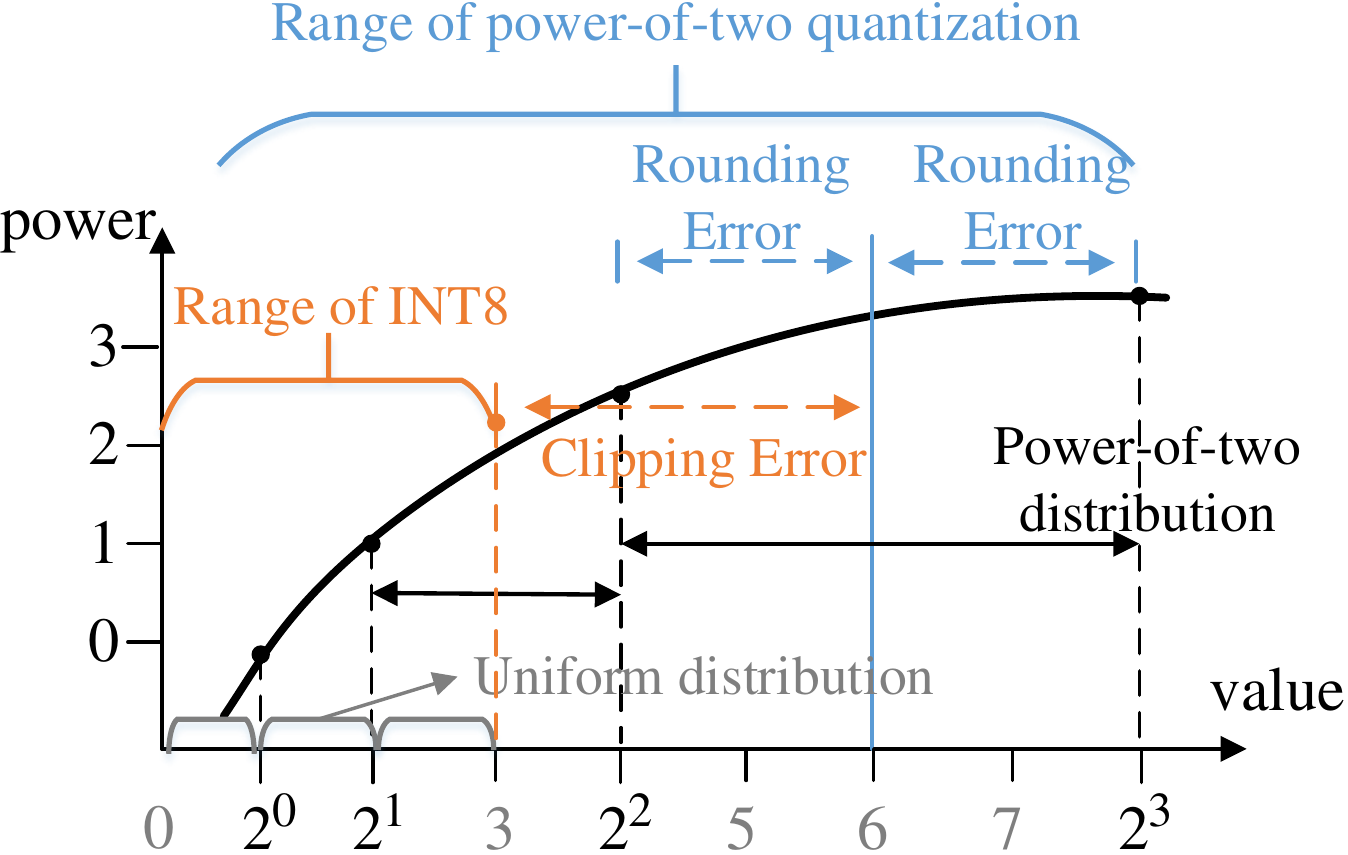}
					\rule[-.5pt]{0pt}{0cm} }
				
			\end{minipage}
		}
		\subfigure[\small An example of double superposition of power-of-two values in AUSN quantization method]{   
			\label{fig:add-example}
			\begin{minipage}[t]{0.45\linewidth}
				\centering  
				\fbox{\rule[-.5pt]{0pt}{4pt} 
					\includegraphics[width=0.9\linewidth]{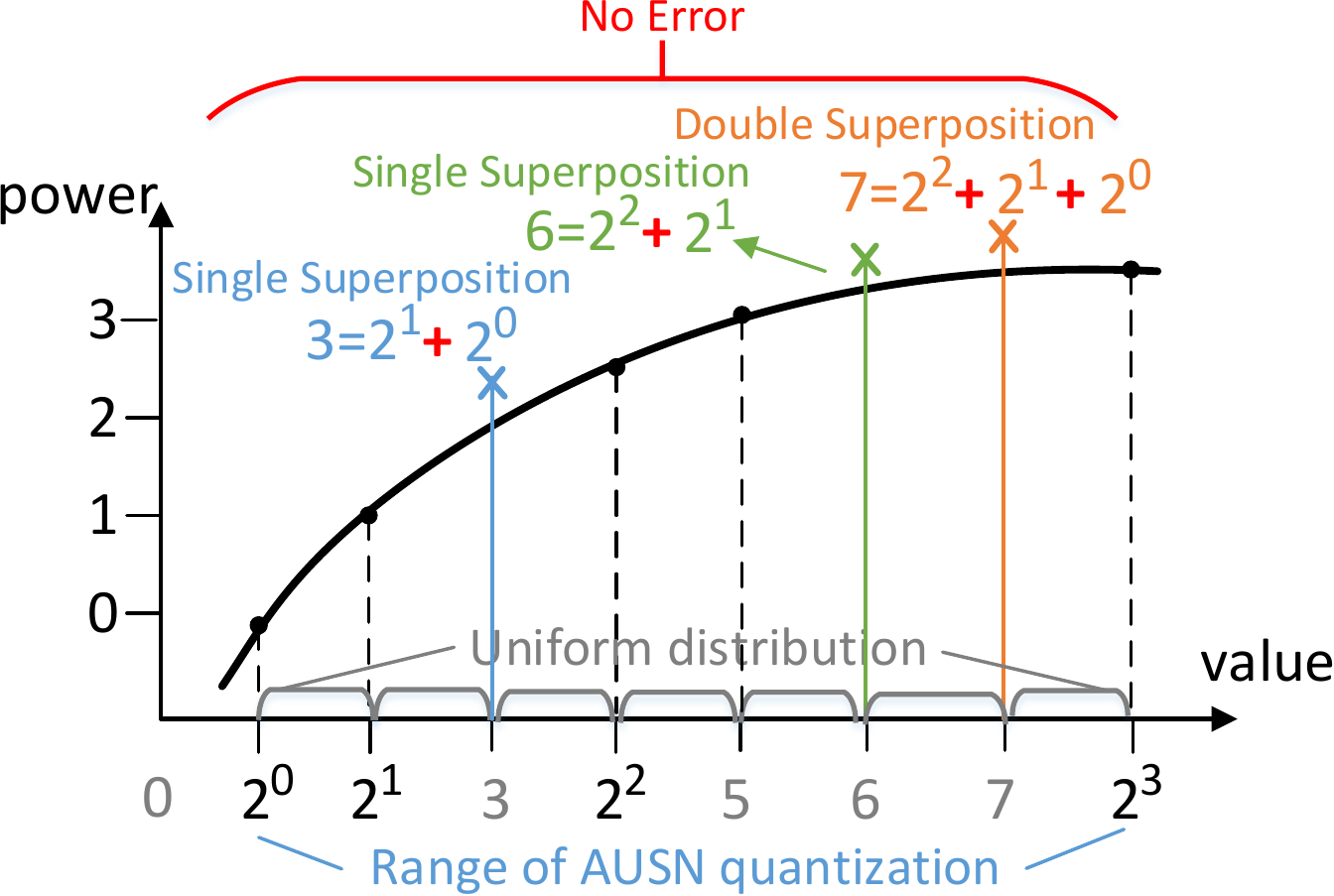}
					\rule[-.5pt]{0pt}{4pt} }
				
			\end{minipage}
		}
		\captionsetup{font={small}}
		\caption{\small The advantages of AUSN over other quantizations: (a) shows the INT8 quantization rounds the weights up/down to the evenly distributed integers. The quantized number in power-of-two quantization is distributed exponentially. Both will cause Clipping Errors and rounding errors, respectively. (b) describes that AUSN eliminates the above errors}
		\vspace{-1em}
	\end{figure}
	Uniform quantization pursues finer \emph{Representing Resolution} result in the larger \textbf{\emph{Clipping Error}} with the limited bit-width. \textcolor{black}{The finer-grain \emph{Representing Resolution} is, the more narrow the \emph{Representing Range} will be, and verse visa. Thus, the accuracy of the DNN model inevitably decreases as the bit-width reduces. As shown in Fig.~\ref{fig:po2-shortcoming}, the values quantized to 2bit by the uniform quantization with the same \emph{Representing Resolution} can only represent $[0, 3]$, if a value of $6$ has to be quantized to the boundary $3$, result in the significant \textbf{\emph{Clipping Error}}. }
	
	In contrast, non-uniform quantization focus on the \emph{Representing Range}, which cause the higher \textbf{\emph{Rounding Error}}. Quantized numbers with 2bit using power-of-two quantization can represent a value in the \emph{Representing Range} of  $[2^0, 2^3]$ in Fig.~\ref{fig:po2-shortcoming}. However, the coarse-grain \emph{Representing Resolution} appears when the exponent is large. In this example, a weight value of $6$ has to be rounded to $2^2$ or $2^3$, rendering significant \textbf{\emph{Rounding Error}}. Also, the power-of-two quantization requires one extra bit to represent the sign of power. 
	
\begin{figure}[t!] 
		\centering  
		\setlength{\belowcaptionskip}{-1.5em}   
		\setlength{\abovecaptionskip}{-0.1em}  
		\subfigure[\small An example for divsion of bit-width]{
			\label{fig:bitdivision}
			\begin{minipage}[t]{0.4\linewidth}
				\centering
				\fbox{\rule[-.5pt]{0pt}{0cm} 
					\includegraphics[width=0.75\linewidth,height=4cm]{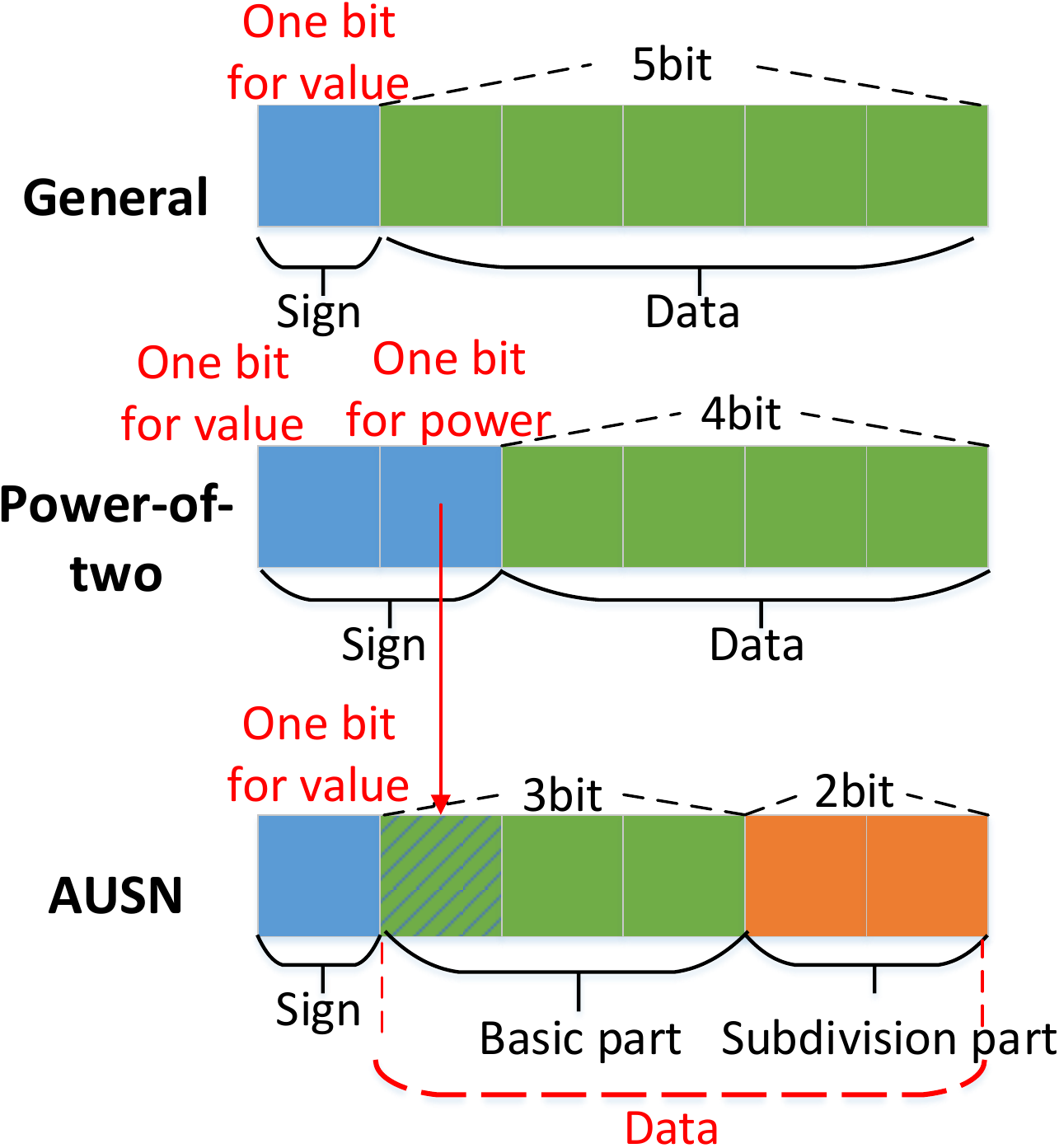}
					\rule[-.5pt]{0pt}{0cm} }
				
			\end{minipage}
		}
		\subfigure[\small The weight distribution of AlexNet model]{   
			\label{fig:ours-alexnet}
			\begin{minipage}[t]{0.55\linewidth}
				\centering  
				\setlength{\fboxsep}{0.1cm}
				\fbox{\rule[-.5pt]{0pt}{0cm} 
					\includegraphics[width=0.9\linewidth,height=4cm]{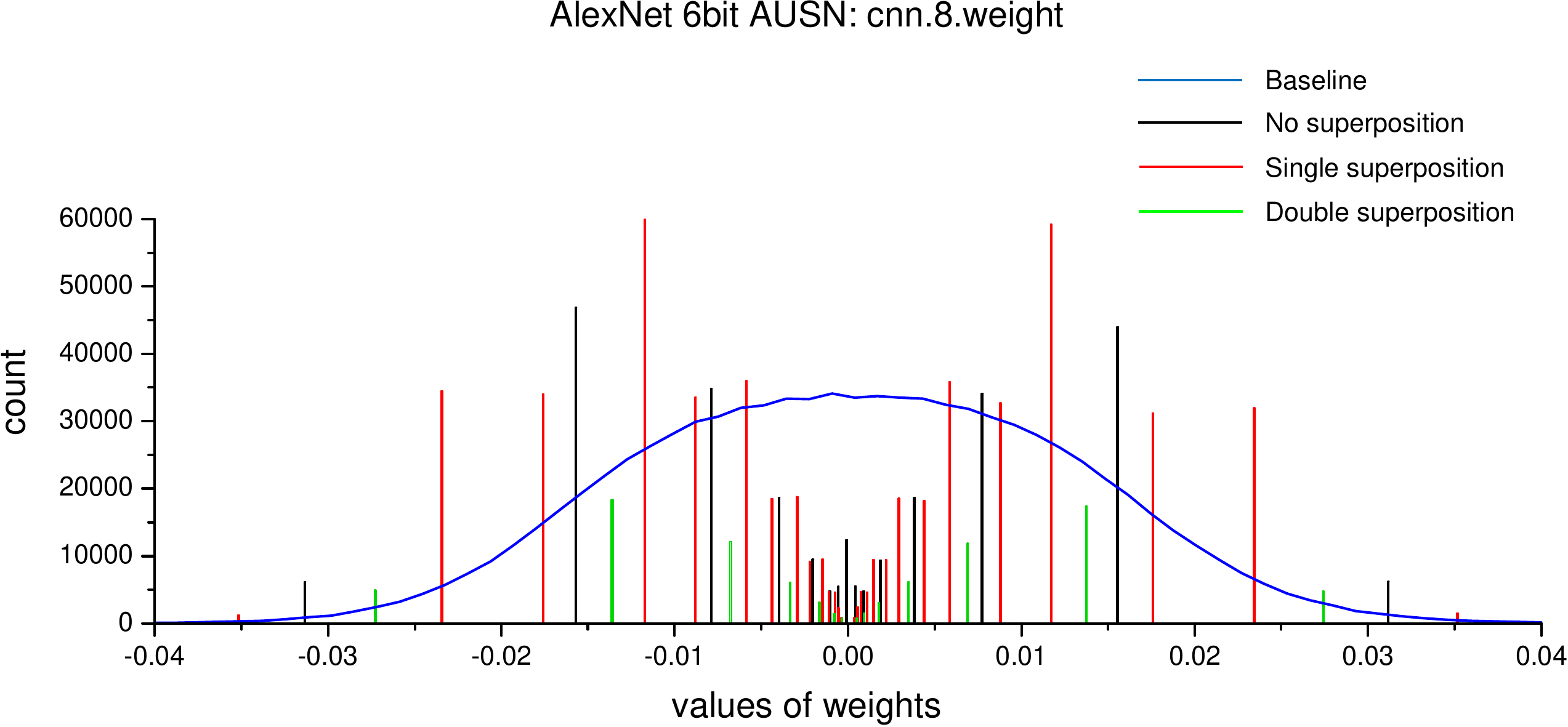}
					\rule[-.5pt]{0pt}{0cm} }
				
			\end{minipage}
		}
		\captionsetup{font={small}}
		\caption{\small The data format and quantized model by AUSN quantization: (a) shows the data format of AUSN quantization. (b) describes the weight distribution of quantized model by AUSN quantization and the original weight distribution of model (the blue line). The black, red and green represent for the values that was not superposed/ superposed once/ superposed twice by power of two's respectively}
	\end{figure}

	For a given bit-width, there is always a trade-off between \emph{Representing Range} and \emph{Representing Resolution}. Our method is also driven by the idea of eliminating these errors. The fundamental objective is to improve the power-of-two quantization to have finer \emph{Representing Resolution} while keeping its good property of having a wide \emph{Representing Range} with low bit-width. 
	\textcolor{black}{Most of the weights of CNN are concentrated near zero and few of them have large absolute value as shown in Fig.~\ref{fig:ours-alexnet}. This phenomenon is well-known as the ``long-term effect'' and is commonly observed in most CNNs. Therefore, it is not worthy of using a large amount of bits to represent all the weights evenly. We find an opportunity to reduce the bit-width of the quantized numbers without accuracy loss.}

	We leverage the superposition of multiple power-of-two quantized numbers to represent the data, rather than rounding the data to the closest power-of-two. As the example in Fig.~\ref{fig:add-example}, we can represent two weight values, e.g., $3$ and $7$, with the superposition of two or three quantized numbers, i.e., $ {3} = {2}^ {1} + {2}^ {0} $ and  $ {7} = {2}^ {2} + {2}^ {1} + {2}^ {0}$, which eliminates the rounding error.

\begin{figure}[h] 
		\centering  
		\vspace{-0.1em}
		\setlength{\belowcaptionskip}{-0.1em}  
		\setlength{\abovecaptionskip}{-0.1em}  
		\fbox{\rule[-.5pt]{0pt}{4pt} 
			\includegraphics[width=0.6\linewidth]{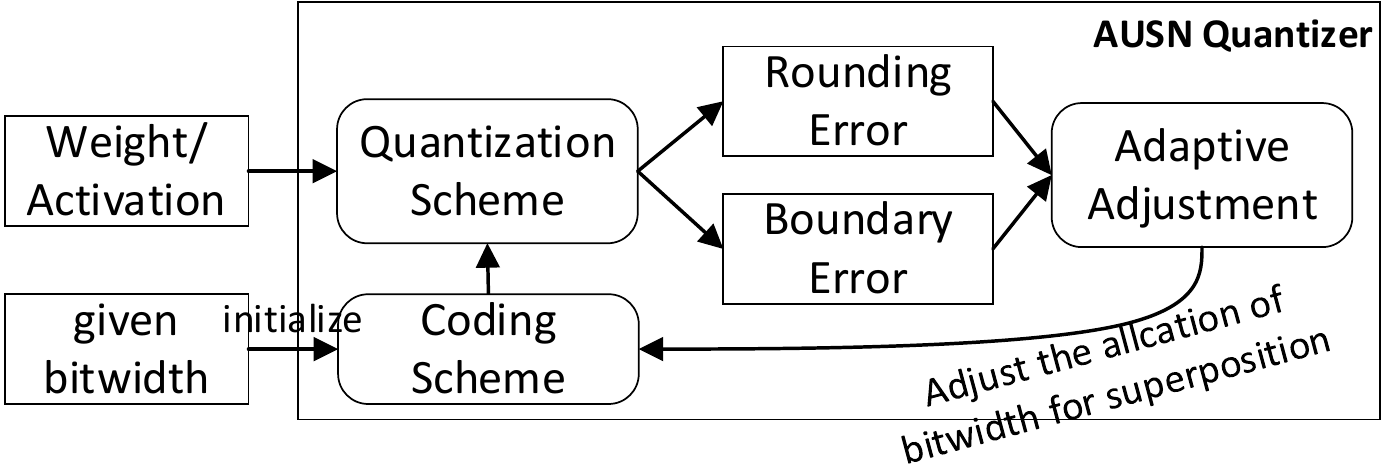}
			\rule[-.5pt]{0pt}{4pt} }
		\captionsetup{font={small}}
		\caption{\small The execution process of AUSN Quantizer}
		\label{fig:quantizer}
	\end{figure}

    \textcolor{black}{However, the weight distribution across the layers of the model is different\cite{wang2019haq}.
	Thus, it is critical to choose the proper number of superposition to make the quantized value ``close" to the original value. More important, Our method (as shown in Fig.~\ref{fig:quantizer}) AUSN can also adaptively allocate the composition of bit-width according to the weight distribution of layers. To combine the averagely high accuracy with finer \emph{Representing Resolution} from uniform quantization and the broad \emph{Representing Range} and hardware acceleration from power-of-two method, we subdivide the limited bit-width and find the balance between \textbf{\emph{Clipping Error}} and \textbf{\emph{Rounding Error}}.}
	\section{AUSN Quantization}
	\subsection{Decoder-free coding scheme for superposition}
	We divide the bit-width into three parts, which are ``sign'',``basic part'', and ``subdivision part''\footnote{The following solutions discuss the data part (unsigned number).}. In Fig.~\ref{fig:bitdivision}, taking 6bit as an example, the bit-width of general quantization consists of one sign bit and five data bits, while bitwidth of power-of-two quantization is composed of two sign bits and four data.  AUSN separates the original five-bit data into the three-bit ``basic'' part and two-bit ``subdivision'' part. By sharing the same sign bit of the value, the superposition of the ``basic'' part and ``subdivision'' part can save 1bit. Given the total bit-width, the proposed AUSN coding scheme allocates the bit-width to the ``basic'' and ``subdivision'' part to ensure a good trade-off between resolution and the range of weight distribution according to the later adaptive adjustment.
		
	\subsection{Quantization scheme with superposition}

	\emph{Note that the value stored in quantized bits is not the quantized number, but the power}. If $ {B}_{basic} $ bits are used to indicate the basic part, we have an intuitive power basis ${Pow}_{ori} = \{0,{2}^{-1},\dotsc, {2}^{-\left({{2}^{{B}_{basic}}-1}\right)}\}$, whose powers are all negative. Thus, we can save them as unsigned numbers and save one ``sign'' bit. In practice, however, the range of ${Pow}_{ori}$ may mismatch with that of ${W}_{j}$. Thus, we introduce a scaling operation called ``PreConvert'' to make the \emph{Representing Range} of the quantized model closer to the weight distribution of original model, and reduce the \textbf{\emph{Clipping Error}}. Since the values stored in bit-width are the powers, the scaling operation on the original values is converted to the shift operation on the power.
	
	\vspace{-0.5em}
	The specific process of AUSN quantization is preformed as follows:
	
	
	First, we calculate the power indicating the largest weight of the layer $ 
	{W}_{j} $:
	\begin{equation}
		\setlength\abovedisplayskip{0pt}
		\setlength\belowdisplayskip{0pt}
		power_{j}=\lceil  \log_2 \max\left( \left|{W}_{j}\right| \right) \rceil
	\end{equation}
	
	Then, we perform the ``PreConvert" operation on ${pow}_{ori}$:
	\begin{equation}
		\setlength\abovedisplayskip{0pt}
		\setlength\belowdisplayskip{0pt}
		{Pow}_{pre} = {Pow}_{ori}*2^{power_j}
	\end{equation}
	
	We denote the power of basic part after ``PreConvert'' operation $Pow_{pre}$, as $\mathcal B_0$, and it satisfies Equation (\ref{con:equ1-3}), which means we use ${Pow}_{pre}$ to approximate the range of ${W}_{j}$. 
	\begin{equation}
	\setlength\abovedisplayskip{3pt}
	\setlength\belowdisplayskip{0pt}
	\max\left( {Pow}_{pre}\right) \leq \max\left( \left|{W}_{j}\right| \right) 
	\leq 2 \max\left( {Pow}_{pre}\right)
	\label{con:equ1-3}
	\end{equation}
	
	Furthermore, for the $n$-th superposition that have $B_1, B_2, \cdots, B_n$ bits respectively, we also have the corresponding superposition power $\mathcal B_i = \{0,2^{-1}, \cdots, 2^{-(2^{B_i}-1)}\}$. Then, we use Algorithm \ref{alg:AUSN-algorithm} to derive the superposition of each quantized number $w_q$.
	\begin{algorithm}[tb]
		\setlength\abovedisplayskip{0pt}
		\setlength\belowdisplayskip{0pt}
		\caption{The procedure for AUSN quantization with Superposition}
		\label{alg:AUSN-algorithm}
		
		\textbf{Input}: {Weight $w$, maximum number of superposition $n$, power basis for each superposition $\mathcal B_i$}\\ 
		\textbf{Output}:  \hspace{-1mm}
		{The quantized number of $w$, $ w_q$}
		
		\begin{algorithmic}[1] 
			\STATE Let $rem\leftarrow w$, current superposition number $tier\leftarrow0$\;
			\WHILE{$rem>0$ {\rm and} $tier\leq n$}
			\STATE  $w_q[tier] = \mathop{\rm argmax} \limits_p \mathcal B_{tier}[p] \leq rem$
			\STATE        $rem = rem/ w_q[tier]$ -1 \;
			
			\STATE        $tier$ += 1 \;
			\ENDWHILE
			\STATE \textbf{return} $w_q=\mathop{{\rm 
					Join}}\limits_{0\leq i\leq n} w_q[\,i\,]$\;
			
		\end{algorithmic}
	\end{algorithm}
	In the procedure of AUSN quantization with superposition defined (as shown in Fig.~\ref{fig:ours-alexnet}), a proximate representation of $w$ can be expressed as 
	\begin{equation}
	\setlength\abovedisplayskip{0pt}
	\setlength\belowdisplayskip{0pt}
	w\doteq\sum\limits_{i=0}^n \prod\limits_{j=0}^i 2^{-w_q[\,j\,]}.
	\end{equation}
	Our AUSN algorithm can naturally round the weight $w$ to the quantized number and decide whether to superpose.
	We can also apply Algorithm \ref{alg:AUSN-algorithm} to the activations. 
	\vspace{-0.2em}
	\subsection{Adaptively adjusting the coding scheme}
	Our AUSN algorithm can naturally round the weight $w$ to the quantized number and decide whether to superpose, and also adaptively adjust internal allocation of bit-width to the base part and the subdivision part according to the \textbf{\emph{Clipping Error}} and \textbf{\emph{Rounding Error}}. For example, AUSN quantization with 5bit, the weight distribution of CONV1 layer, the 4bit for the basic part and 1bit for the subdivision part, while 3bit for the basic part and 2bit for the subdivision part in CONV2 layer. 
	Suppose that the initial distribution of the weights in a layer is normal. Without losing universality, we suppose that the weight follows standard normal distribution $\mathcal N(0,1)$ and is clipped at $(-\mathcal L, \mathcal L)$. Since quantization distribution is usually evenly distributed about 0, we only consider the positive part. Suppose that AUSN quantization has a \emph{Representing Range} $[0, R]$ and a set of quantized numbers $\mathcal P$, we can use the following equation to present the \textbf{\emph{Clipping Error}} $\mathbb E_b$ and \textbf{\emph{Rounding Error}} $\mathbb E_r$:
    \begin{align}
    \vspace{-2em}
    \setlength\abovedisplayskip{0pt}
	\setlength\belowdisplayskip{0pt}
    \label{con:equ41}
    	\mathbb E_b = \int_R^\mathcal L w \cdot |w - R| \cdot \mathcal N(w;0,1)\ dw
        \qquad	
    	\mathbb E_r = \int_0^R w \cdot \min\limits_{p\in\mathcal P} |p-w| \cdot \mathcal N(w;0,1) \ dw
    \end{align}
    First, AUSN initializes the coding scheme according to the given bit-width and quantizes the weights. Then, the \textbf{\emph{Clipping Error}} $\mathbb E_b$ and \textbf{\emph{Rounding Error}} $\mathbb E_r$ between the quantized weights and original weights is calculated according to euqation(\ref{con:equ41}). Last, based on $\mathbb E_r$ and $\mathbb E_b$, AUSN adjusts the boundary $\mathcal R$ of \emph{Representing Range} and the set of quantized value $\mathcal P$(that is, the allocation of bit-width). The above process is iterated until an optimal allocation of bit-width is found.

	
	\subsection{AUSN rounding scheme}
		
		Although AUSN quantization has effectively solved the problem of the 
		power-of-two quantization in the reduction of accuracy, there still 
		exists two critical issues for almost quantization methods:
		\begin{enumerate}
		    \vspace{-0.5em}
		    \setlength{\itemsep}{0pt}
			\item \emph{Overflow of Bit-width:} The output resulted from the 
			convolution operation (Multiply and Accumulate) occupies a larger 
			bit-width to maintain the precision than that of the quantized 
			number. For example, the convolution operation of two $2\times 2$ 
			matrix consisting of all 8-bit numbers first result in four 16-bit 
			dot-products and then accumulate to the 32-bit sum.
			\item \emph{Overhead of Re-quantization:} In deep neural networks, 
			the output of the previous layer is the input of the next layer. 
			The outputs have a high likelihood of exceeding 
			the range of the quantized numbers for the input layer. Thus, 
			existing methods have to re-quantize the output values — such 
			re-quantization results in significant cost of computation and 
			resource.
		\end{enumerate}
		We formulate the above problems as follows. Suppose the next 
		layer only accepts an activation value with a double superposition due 
		to the bit-width limit, but in the process of multiplication with AUSN, the number of superposition will increase, e.g., $(a+b)\times(c+d) = A+B+C+D$.Thus, we have to eliminate some power-of-two 
		terms (one term in this example) and round up/down the activation value 
		with a minimized rounding error. Consequently, after the multiplication 
		operation, we can substitute the accumulation operation with rounding 
		scheme, which eliminates the overflow and re-quantization issue.

		The basic principle of rounding operation is simple enough and to 
		minimize the rounding error. Given the data in the form of a polynomial 
		of power-of-two, we need to determine the power after rounding.
		{\bf Principle:} ${2}^{m} \leq data \leq {2}^{m+1} $, ${m} \in 
		\mathbb{N}$, so that $Quant\left(data\right) \approx {2}^{m} \,$ or $\, 
		Quant\left(data\right)\approx {2}^{m+1}$.
		
		The rounding scheme can be categorized into the following four 
		scenarios:
		
		{\bf Scenario 1:}$\quad \underbrace{{2}^{n}+{2}^{n+1}+\cdots+{2}^{m} 
		}_{m-n+1 \geq {B}_{sub} +2}\approx {2}^{m+1}\,$, wherein ${B}_{sub}$ is 
		the bit-width of the subdivision part indicating the number of 
		superposition.		
		
		%
		
		
		{\bf Scenario 2:}$\quad$When the condition ${m}={n}$ is satisfied, it's 
		obvious that $\,{2}^{n}+{2}^{m}\approx{2}^{m+1}$.
		
		{\bf Scenario 3:}$\quad$If ${m} > {n} + {B}_{sub}$, then 
		${2}^{n}+{2}^{m}\approx{2}^{m}$.
		
		{\bf Scenario 4:}$\quad$If ${m} = {n} + {B}_{sub}$, then 
		${2}^{n}+{2}^{m}\approx{2}^{m}$ or ${2}^{n}+{2}^{m}\approx{2}^{m+1}$.
		

		
		Based on the above scenarios, we round the data using the following 
		algorithms. 
		\begin{itemize}
		    \vspace{-1em}
			\setlength{\itemsep}{0pt}
			\setlength{\parsep}{0pt}
            \setlength{\parskip}{0pt}
			\item[Step 1:] Find the maximum terms in data that can apply 
			Scenario 1.
			\item[Step 2:] Find pair of terms that can apply Scenario 2.
			\item[Step 3:] Apply Scenario 3 if the condition 
			$num\left(rest\right) \geq {B}_{sub} + 2$ is met. 
			\item[Step 4:] If Equation (\ref{con:equ5}) work, carry out the 
			Scenario 4.
		\end{itemize}
		\begin{equation}
			\setlength\abovedisplayskip{-1pt}
			\setlength\belowdisplayskip{-3pt}
			\begin{cases}
			num\left(rest\right) = {B}_{sub} + 1\\
			{B}_{sub} < bit-width\left(min\left(rest\right)\right)
			\end{cases}
			\label{con:equ5}
		\end{equation}
		Following the above steps, we can eliminate the number of power-of-two 
		to satisfy ${B}_{sub}$ and minimize the rounding error. Take the data 
		$412 = 2^2+2^3+2^4+2^6+2^6+2^8$ as example, $\text{assuming}\,{B}_{sub} 
		= {1}$, that is the quantized number is single superposition of 
		power-of-two. Step1: merge the $2^5 = 2^2+2^3+2^4$ and get the 
		$2^5+2^6+2^6+2^8$; Step 2:merge the $2^7 = 2^6+2^6$ and get the 
		$2^5+2^7+2^8$; Then, the condition of step 3 is not met, and thus step 
		4 is performed: merge the $2^7 = 2^5 + 2^7$ and get the $2^7+2^8$; 
		Finally, we find the result $384=2^7+2^8$ satisfy the requirement of 
		quantization.

	
	

	

	\section{Experiments}
	To evaluate the performance of our algorithm, We quantize several models 
	and verify the accuracy of the models on the  different datasets. By using 
	Cifar10~\cite{krizhevsky2009learning} and ImageNet~\cite{deng2009imagenet}, 
	we compare AUSN with several state-of-the-art quantization methods on CNN. 
	We implement the AUSN quantization using the CNN model structures and 
	pre-trained models in Pytorch 
	library\footnote{\url{https://github.com/pytorch/vision/tree/master/torchvision/models}}. 

	
	In addition, we verified the sequential network and the network used in 
	target detection task through the Google Speech Commands dataset, the 
	VoxCeleb dataset and COCO datasets. We compared our results with full-precision baselines models to demonstrate the broad applicability of our approach.
	
	Remarkably, we quantize the pre-trained models without fine-tuning or retraining.
	\subsection{Result of quantized networks on Cifar10}

	From the data in table \ref{tab:cifar_result}, we can see that both the 
	self-adapting shifting and the better resolution brought by subdivision 
	part enhenced our result. For the 2-bit result, both schemes quantize 
	weights into power-of-twos, but our AUSN find a better covering range and 
	thus get a better result. For the result of 3, 4 and 5-bit quantization, 
	our AUSN method retained the basic part as 3 bits and set the subdivision 
	part as 0, 1 and 2 bits, respectively. Compared with the INQ result, we can 
	find that our accuracy progress for each bit added is much larger than that 
	of INQ counterpart. This means that when the representation range is 
	guaranteed (presented by the bit-width for the basic part in AUSN or the 
	whole bit-width in INQ), it is the resolution (presented by the bit-width 
	for the subdivision part in AUSN) that helps the accuracy to grow. 
	
	Especially, we quantized the pruned model of ResNet-18 to show the 
	harmonious cooperation between the AUSN and the pruning method.

	\newcommand{\tabincell}[2]{\begin{tabular}{@{}#1@{}}#2\end{tabular}}
	\renewcommand{\arraystretch}{1.2} 
	\setlength{\tabcolsep}{0.5mm}{
		\begin{table*}[htbp]  
			\vspace{-0.8em}
			\setlength\abovedisplayskip{-3pt}
			\setlength\belowdisplayskip{-3pt}
			\centering 
			\fontsize{6.5}{8}\selectfont  
			\captionsetup{font={small}}
			\caption{\small Accuracy on Cifar10 for typical quantized networks 
				comparison by different bit-width. The last one is the pruned 
				ResNet50 model at 50\% sparsity. Compared quantization is INQ .}
			\label{tab:cifar_result}  
			\begin{tabular}{cccccccccc}
				
				\toprule[1pt]
				
				\multirow{2}{*}{Model}&\multirow{2}{*}{\tabincell{c}{Baseline\\32bit}}&
				\multicolumn{4}{c}{AUSN quantization 
					(Ours)}&\multicolumn{4}{c}{power-of-two(INQ 
					~\cite{zhou2017incremental})}\\
				\cmidrule(lr){3-6}\cmidrule(lr){7-10}
				~ & ~ & 5bit/Top-1(\%) & 
				4bit/Top-1(\%) & 3bit/Top-1(\%) & 2bit/Top-1(\%) &
				5bit/Top-1(\%) & 4bit/Top-1(\%) & 3bit/Top-1(\%) & 
				2bit/Top-1(\%)   \\
				
				\midrule
				AlexNet &85.13  & \color{red}\textbf{85.21} & 85.00 & 84.80  
				&83.21 
				& 81.90    &81.89          &80.29          &80.27       \\ 
				
				GoogleNet &90.89  & \color{red}\textbf{90.85}    & 90.01 & 
				89.37  &89.11 
				& 88.28    &88.28          &88.27          &88.21       \\ 
				
				VGG16 &93.92  & \color{red}\textbf{93.95}    & 93.58 & 93.68  & 
				92.75 
				& 91.88    &91.84          &91.27          &91.26       \\ 
				
				ResNet18  &92.22  & \color{red}\textbf{92.25} & 91.83 & 91.80  
				&91.46 
				& 90.47    &90.31          &90.14          &90.10       \\ 
				
				ResNet18(pruned)  &94.17  & \color{red}\textbf{94.11}    & 
				94.09 & 93.52  &92.65 
				& 92.87    &92.85          &92.83          &88.15       \\    
				\toprule[1pt]
			\end{tabular}
			\vspace{-1em}
		\end{table*}  
	}
	
	\subsection{Result of quantized networks on ImageNet}
	
	In the experiments on ImageNet, we choose the most widely used models 
	ResNet18 and ResNet50 and compare our result with other quantization 
	schemes. Table~\ref{tab:comparsion} shows that the Top-1 accuracy loss in 
	our 5-bit AUSN is less than 0.1\%, far better than other schemes. 

    \begin{wraptable}{r}{6.5cm}
		\centering
 		\vspace{-0.3cm}
		\setlength\abovedisplayskip{-3pt}
		\setlength\belowdisplayskip{-1pt}
		\fontsize{8}{10}\selectfont  
		\captionsetup{font={small}}
		\setlength{\tabcolsep}{0.5mm}{
		\caption{\small ResNet18 on ImageNet. 
			Top-1 accuracy loss (\%) with quantized weights and activations by 
			various quantization method, including SR+DR~\cite{gysel2018ristretto}, INT~\cite{jacob2018quantization}, RQ ST~\cite{louizos2018relaxed}, PACT~\cite{choi2018pact}, QIL~\cite{jung2019learning}.}
		\centering
		\label{Tab:ImageNet_both}
		\begin{tabular}{ccccccc}
			\toprule
			Method   & SR+DR& INT   & RQ ST & PACT& QIL & \textbf{Ours}  \\ 
			\midrule
			Bit-width   & 5/5     & 5/5 &  5/5  
			& 5/5    &5/5 & \textbf{5/5}  \\ 
			Acc. loss        & -10.5 & -2.5 & -1.6 & -0.3 & -0.8 &\color{red}\textbf{-0.3}\\
			
			\toprule
		\end{tabular}
		}
		
		\vspace{-1cm}
	\end{wraptable}
	Meanwhile, we quantize the weights of pre-trained models without fine-tuning 
	and got the results in Table \ref{tab:comparsion}, which means that the 
	time of quantization has significantly been saved. Also, in Table~\ref{Tab:ImageNet_both}, we quantize both weights and activations  to 
	optimize the inference process. According to the advantages mentioned 
	above, our AUSN method reached state-of-the-art result in existing 
	quantization schemes. 
	
	\subsection{Result of quantizing sequential models and Yolo}
	We explore the effect of the AUSN quantization on the sequential network such as GRU, CRNN~\cite{zhang2017hello}, and TDNN~\cite{snyder2018x} through the Google Speech Commands dataset and the VoxCeleb dataset, and the effect of the AUSN quantization on the target detection task such as YOLOv3-tiny through the COCO datasets, as shown in Table \ref{tab:comparsion}. There is few 
	decrease in accuracy  (\textit{i}.\textit{e}. {mAP} for Yolov3-tiny in COCO 
	and Top-1 accuracy for sequential models) for quantizing weights from 32bit 
	to 4bit or 5bit.

	\begin{table}[t!] 
		\renewcommand{\arraystretch}{1.1}
		\small
		\captionsetup{font={small}}
		\caption{\small The accuracy comparison of DNNs uses existing methods on the datasets of image recognition, target detection, and speech recognition tasks. The quantized models by AUSN quantization without fine-tuned and any augmentation tricks.} 
		
		\label{tab:comparsion}
		\resizebox{\textwidth}{!}{
			\begin{threeparttable}
				\centering
				
				\begin{tabular}{cccc||cccc}
					\toprule
					\multicolumn{8}{c}{ImageNet Dataset}\\
					\toprule
					Method & bit-width & Top-1(\%) & Drop in Top-1(\%) &Method & 
					bit-width & Top-1(\%) & Drop in Top-1(\%)\\
					\midrule
					\multicolumn{4}{c||}{ResNet50* (baseline: 76.13\%)} & 
					\multicolumn{4}{c}{ResNet18* (baseline: 69.76\%)}\\
					\midrule
					Dual ~\cite{choukroun2019low}& 4bit & 70.20 & -5.93 & BWN ~\cite{rastegari2016xnor} & 2bit & 60.8 & -8.5\\
					INQ~\cite{zhou2017incremental} & 5bit & 74.81 & -1.59 & Dual ~\cite{choukroun2019low}& 4bit &  66.63 & -3.13\\
					Focused compression ~\cite{zhao2019focused}& 5bit & 74.86 & -1.54 & LAPQ~\cite{nahshan2019loss} & 4bit & 
					62.6 & -7.16\\
					ADMM Quantization ~\cite{ye2019progressive}& 6bit & 75.93 & -0.2 & Focused compression  ~\cite{zhao2019focused}& 5bit 
					&  68.36 & -1.40\\
					SYMM ~\cite{nayak2019bit}& 6bit & 72.58 & -3.55 & UNIQ ~\cite{baskin2018uniq} & 5bit &  68.00 & 
					-1.76\\
					Biscaled-FxP ~\cite{jain2019biscaled}& 6bit & 70.46 & -5.67 & DFQ ~\cite{nagel2019data} & 6bit &  66.3 & -3.4\\
					V-Q ~\cite{park2018value}& 7bit & 75.89 & -0.24 &INT8  ~\cite{jacob2018quantization}& 8bit &  67.3 & -2.4\\
					INT8 ~\cite{jacob2018quantization}& 8bit & 74.9 & -1.5 & RQ ~\cite{louizos2018relaxed}& 6bit &  68.6 & -1.16\\
					\toprule[0.85pt]
					
					\textbf{Ours} & 4bit & \textbf{75.37} & \textbf{-0.76} & 
					\textbf{Ours} & 4bit & \textbf{68.84} & \textbf{-0.92} \\
					\rowcolor[HTML]{FFD19A}
					\textbf{Ours} & 5bit & \textbf{76.09} & 
					\color{red}\textbf{-0.04} & \textbf{Ours} & 5bit & 
					\textbf{69.67} & \color{red}\textbf{-0.09} \\
					\bottomrule
					\toprule
					\multicolumn{8}{c}{Speech Command Dataset}\\
					\toprule
					Method & bit-width & Top-1(\%) & Drop in Top-1(\%) &Method & 
					bit-width & Top-1(\%) & Drop in Top-1(\%)\\
					\midrule
					\multicolumn{4}{c||}{GRU (Network cfg\tnote{$\dagger$}\;: S = 
						40, N = 154)} & \multicolumn{4}{c}{GRU (Network cfg\tnote{$\dagger$}\;: S = 
						20, N = 400)}\\
					\midrule
					baseline   & 32bit & 93.62 & - & baseline & 32bit & 94.62 & 
					-       \\
					\toprule[0.85pt]
					
					\textbf{Ours} & 4bit & \textbf{93.15} & \textbf{-0.47} & 
					\textbf{Ours} & 4bit & \textbf{94.34} & \textbf{-0.28} \\
					\rowcolor[HTML]{FFD19A}
					\textbf{Ours} & 5bit & \textbf{93.47} & 
					\color{red}\textbf{-0.15} & \textbf{Ours} & 5bit & 
					\textbf{94.57} & \color{red}\textbf{-0.05} \\
					\bottomrule

					\toprule
					\multicolumn{8}{c}{Speech Command Dataset}\\
					\toprule
					Method & bit-width & Top-1(\%) & Drop in Top-1(\%) &Method & 
					bit-width & Top-1(\%) & Drop in Top-1(\%)\\
					\midrule
					\multicolumn{4}{c||}{CRNN$_A$ \tnote{$\ddagger$}} & 
					\multicolumn{4}{c}{CRNN$_C$ \tnote{$\ddagger$}}\\
					\midrule
					baseline   & 32bit & 93.50 & - & baseline & 32bit & 94.56 & 
					-       \\
					\toprule[0.85pt]
					
					\textbf{Ours} & 4bit & \textbf{93.17} & \textbf{-0.33} & 
					\textbf{Ours} & 4bit & \textbf{94.28} & \textbf{-0.28} \\
					\rowcolor[HTML]{FFD19A}
					\textbf{Ours} & 5bit & \textbf{93.38} & 
					\color{red}\textbf{-0.12} & \textbf{Ours} & 5bit & 
					\textbf{94.48} & \color{red}\textbf{-0.08} \\
					\bottomrule

					\toprule
					\multicolumn{4}{c}{VoxCeleb 
						Dataset}&\multicolumn{4}{c}{COCO datasets}\\
					\toprule
					Method & bit-width & Top-1(\%) & Drop in Top-1(\%) &Method & 
					bit-width & mAP & Drop in Acc.\\
					\midrule
					\multicolumn{4}{c||}{TDNN} & 
					\multicolumn{4}{c}{YOLOv3$-$tiny}\\
					\midrule
					baseline   & 32bit & 80.38 & - & baseline & 32bit & 33.1 & 
					-       \\
					\toprule[0.85pt]
					
					\textbf{Ours} & 4bit & \textbf{79.98} & \textbf{-0.40} & 
					\textbf{Ours} & 4bit & \textbf{31.6} & \textbf{-1.5} \\
					\rowcolor[HTML]{FFD19A}
					\textbf{Ours} & 5bit & \textbf{80.25} & 
					\color{red}\textbf{-0.13} & \textbf{Ours} & 5bit & 
					\textbf{32.2} & \color{red}\textbf{-0.9} \\
					\bottomrule
				\end{tabular}
				
				\begin{tablenotes}
					\footnotesize
					\item[$\dagger$] {S: Frame Stride, N: Cells of GRU}
					\item[$\ddagger$] {The Network cfg of CRNN$_A$ is: 
						C(48,10,4,2,2)-N(60)-N(60)-F(84); the Network cfg of 
						CRNN$_C$ is: C(100,10,4,2,1)-N(136)-N(136)-F(188), where C: 
						Shape of Convolutional layer, N: Cells of GRU, F: Cells of 
						fully connected layer} 
				\end{tablenotes}
			\end{threeparttable}
		}
	\end{table}

	\subsection{Result of quantized networks deployed on FPGA}

\begin{table}
			\centering
			\fontsize{8}{10}\selectfont  
			\captionsetup{font={small}}
			\caption{\small The comparison of FPGA resources and energy 
				consumption for 64 x 64 MAC array}
			\label{tab:HLS}
			\begin{threeparttable}
			\centering
			\begin{tabular}{ccccccccccc}
				\hline
				\multirow{2}*{Resources} & 
				\multicolumn{3}{c}{\multirow{2}*{\makecell[c]{Multiplier \& \\ Accumulator~\cite{mcdanel2019full}}}} & 
				\multicolumn{6}{c}{Shifter\&Adder} & \multirow{2}*{Performance} 
				\\
				\cline{5-10}
				~ & ~& ~ &~& \multicolumn{3}{c}{need decoder} & \multicolumn{3}{c}{\textbf{Ours}} & ~  \\
				\hline
				~ & {$8/8$}\tnote{1} \qquad& {$8/5$} \qquad& {$8/4$}\qquad& {$8/8$} \qquad&{$8/5$} \qquad& {$8/4$}\qquad&{$8/8$}\qquad&{$8/5$} \qquad&{$8/4$}& ~  \\
				\hline
				LUT     & 212388& 187262  & 181248 & 225280
 & 212942 & 203712 & 133120 & 112071 & 108544 & 2.0$\times$      \\
				FF     & 192293& 143142  & 108729 & 86317 & 512731 & 45729  & 54313  & 45127 & 44032 & 4.4$\times$  \\
				Energy Consumption     &4.21W &3.75W &3.67W &4.51W &4.26W &  4.07W  &2.65W  & 2.24W & 2.17W &  1.9$\times$  \\ 
				\hline
			\end{tabular}
			\begin{tablenotes}
                \item[1] A/B refers to A-bit input multiplied by B-bit weights
           \end{tablenotes}
          \end{threeparttable}
\end{table}
	AUSN doesn't need a decoder. Decoder is necessary only when the control 
	information is encoded into bit-word. AUSN directly ``multiply'' two 
	exponents composing the weight with the input in parallel by shift 
	operations, whose results is added up (superposition) to the output. AUSN 
	is not only as easy to compute as uniform quantization (like INT8), but 
	also hardware friendly because the expensive multiplication operation is 
	replaced by the simple bit shift and add operations. Compared with the 
	traditional 8-bit multiply-accumulator, the shifter can be realized only by 
	logic units such as LUTs. As shown in Table~\ref{tab:HLS}, we found that the quantized model by AUSN 
	not only DSPs are not required, but also saves the LUTs 
	2.0$\times$, FFs 4.4$\times$ and power 1.9$\times$ by using the Xilinx 
	Vivado HLS suite, which dramatically reduces the hardware cost (The evaluation platform is Xilinx ZCU104).
	
	\section{Conclusion}
	In this paper, we introduced a novel quantization method called AUSN, which combines the advantages of broad representing range of power-of-two quantization and the constant representing resolution of the uniform quantization. This method can quantize both the activation and weight of a pre-trained model to low bit-width (less than 8bit) without accuracy loss. AUSN shows an excellent generality because it can adaptively adjust the number of superposition and the coding scheme given a fixed bit-width. We further design a rounding scheme in AUSN to eliminate the overhead of re-quantization. The thorough experiments show the superiority of AUSN quantization against state-of-the-art methods. Compared with other quantization methods, AUSN quantize the pre-trained model without retraining. Notably, the AUSN quantization is hardware-friendly. The synthesis~(see Appendix B for detail) results on FPGA proved that AUSN can effectively reduce the resources and energy consumption. We measure the effectiveness of quantization methods by information loss from the theoretical perspective and analysis of the relationship between the quantization methods and performance of hardware (as in Appendix A). 
	
	\clearpage
	\section*{Broader Impact}
	The high accuracy of DNNs is achieved by consuming a lot of computing and storage resources, which significantly hinders the application of DNNs in edge devices. Quantization is proposed and widely used for storage and computation reduction. As for AUSN quantization, results in fewer bits representing these operands and corresponding operations by lowering the precision demands of operations and operand, which reduces the cost of datapath and memory. Furthermore, it can also reduce the energy consumption and resources of hardware and expand the accelerator design space of many terminal devices with limited hardware resources. It is worth noting that AUSN quantization is a general quantization method, which can be used in many tasks, such as image recognition, speech recognition, target detection.
	
    \printbibliography

	\clearpage
	\appendix
	\begin{appendix}
	\section{Theoretical analysis}
	We explore the choice of quantization in the context of information theory, 
	which enables us to reinterpret quantization as the Minimum Description 
	Length (MDL) problem~\cite{hinton1993keeping,ullrich2017soft}. Both MDL and 
	quantization can be regarded as a search problem, aiming at finding out 
	the optimal method to compress data with balancing the accuracy and the 
	complexity (including bit width, quantization method, etc.) of the model. 
	We define the loss function to measure the loss of information 
	between the distribution of original weights, $M$ and the distribution of quantized weights, $M_q$, mainly 
	including the error and the complexity. Specifically:
	\vspace{-0.3cm}
	\begin{equation}
		\begin{aligned}
		\small
		\vspace{-0.8cm}
			\setlength\abovedisplayskip{-1.5em}
			\setlength\belowdisplayskip{-1.5em}
			\mathcal{L} \left({M}, M_q\right) 
            &= \underbrace{\log\frac {p(D|M)} {p(D|M_q)}}_{\rm the \ error} + 
			\underbrace{\mathbb{KL}(M, M_q)}_{\rm the \ complexity}
			\label{con:equa7}
		\end{aligned}
		\vspace{-0.8em}
	\end{equation}
	
	where ${D}$ denotes the dataset that ${M}$ and ${M_q}$ is trained on. ${-\log p(D|M)/p(D|M_q)}$ indicates the error caused by the misfit between	model ${M}$ and quantized model $M_q$ trained on the dataset $D$. The dataset $D$ consists of inputs ${X}$ and expected output 
	${Y}$, ${p(D|M)}$ can be further expressed as ${p(Y|X, M)}$, which means the
	probability of achieving the correct output ${Y}$ given the input ${X}$ and 
	model ${M}$. ${\mathbb{KL}(M, M_q)}$ indicates the complexity of the quantization methods.


	According to the information theory, ${\mathbb{KL}(P, Q)}$ is used to 
	measure the average number of extra bits required by ${Q}$ encoding ${P}$, 
	i.e., the distance between them. We then consider ${Q}$ as an approximation 
	of the distribution of ${P}$. Thus, ${\mathbb{KL}(M, M_q)}$ represents 
	KL divergence, which measures the similarity between the distributions of 
	${M}$ and $M_q$. 
	For ${\mathbb{KL}(M, M_q)}$, firstly we divide the 
	original distribution of ${M}$ according to the set of points of the 
	quantized distribution $\widehat{\mathbb{M}}$, and then calculate the 
	probability of differences between ${M}$ and $\widehat{\mathbb{M}}$:
	\begin{equation}
	\small
	\setlength\abovedisplayskip{1pt}
	\setlength\belowdisplayskip{1pt}
	\centering
	\begin{aligned}
	{\mathbb{KL}(M, \widehat{\mathbb{M}})} &=  \sum M\left ( X \right 
	)\log\frac{M\left ( X \right )}{\widehat{\mathbb{M}}\left ( Y \right )}\\
	\mathrm{ s.t.\quad}{M\left ( X \right )} = \displaystyle 
	\frac{{sum}_{x}\left({y}_{i}\right)}{\sum_{s\in Y }sum_{x}\left ( s \right 
		)} &\quad
	{sum}_{x}\left({y}_{i}\right) =  
	\displaystyle\int_{\sqrt{{y}_{i-1}{y}_{i}}}^{\sqrt{{y}_{i}{y}_{i+1}}}distri_x(s)\mathrm{d}
	x 
	\end{aligned}
	\label{con:equ8}
	\vspace{-0.2cm}
	\end{equation}
	
	where ${distri}_{x}$ is the distribution of weight in real number field. 
	${sum}_{x}\left(y\right)$ is the approximate count of weights according to 
	${distri}_{x}$ at $y$. $Y$ is the set of quantized numbers.

	\begin{table}[!h]
		\centering
		\renewcommand{\arraystretch}{1.2} 
		\fontsize{8}{10}\selectfont  
		\vspace{-0.32cm} 
		\setlength{\abovecaptionskip}{-0.02cm}   
		\captionsetup{font={small}}
		\caption{The information loss of the AlexNet quantized by AUSN on Cifar10}
		\label{Tab:03}
		\begin{tabular}{ccccc}
			
			\toprule[1pt]
			
			
			\tabincell{c}{Bit-Width} \qquad& Quant \qquad&\tabincell{c}{KL divergence} \qquad& 
			\tabincell{c}{Accuracy loss} \qquad& \tabincell{c}{Information loss}\\
			\toprule[1pt]
			
	        \multirow{2}{*}{5bit}  &Ours &0.014 & -0.001 & \color{red} 
			\textbf{0.013$(1)$}\\
			
			\cline{2-5}
			
			& INQ & 0.004  & 3.23 & 3.234$(4)$\\
			\toprule[1pt]
			
			\multirow{2}{*}{4bit}  &Ours &0.062 & 0.13 & 0.192$(2)$\\
			
			\cline{2-5}
			& INQ & 0.004 & 3.24 & 3.244$(5)$ \\
			\toprule[1pt]
			
			\multirow{2}{*}{3bit}  &Ours &0.261 & 0.15 & 0.411$(3)$\\
			
			\cline{2-5}
			
			& INQ & 0.004  & 4.84 & 4.844$(6)$\\
			
			
			\toprule[1pt]
		\end{tabular}
		\label{Tab:info_loss}
		\vspace{-0.3cm}
	\end{table}

	The smaller value of information loss, the less accuracy loss  induced by quantization. In Table \ref{Tab:info_loss}, the figure in brackets is in 
	the ascending order of information loss. The red label is the best 
	quantization effect. 

		\section{Efficiency Analysis}
		\subsection{Analysis on the Evolution}
\begin{figure}[h]
			\begin{center}
				\fbox{\rule[-.5pt]{0pt}{0cm} 
					\includegraphics[width=0.95\linewidth]{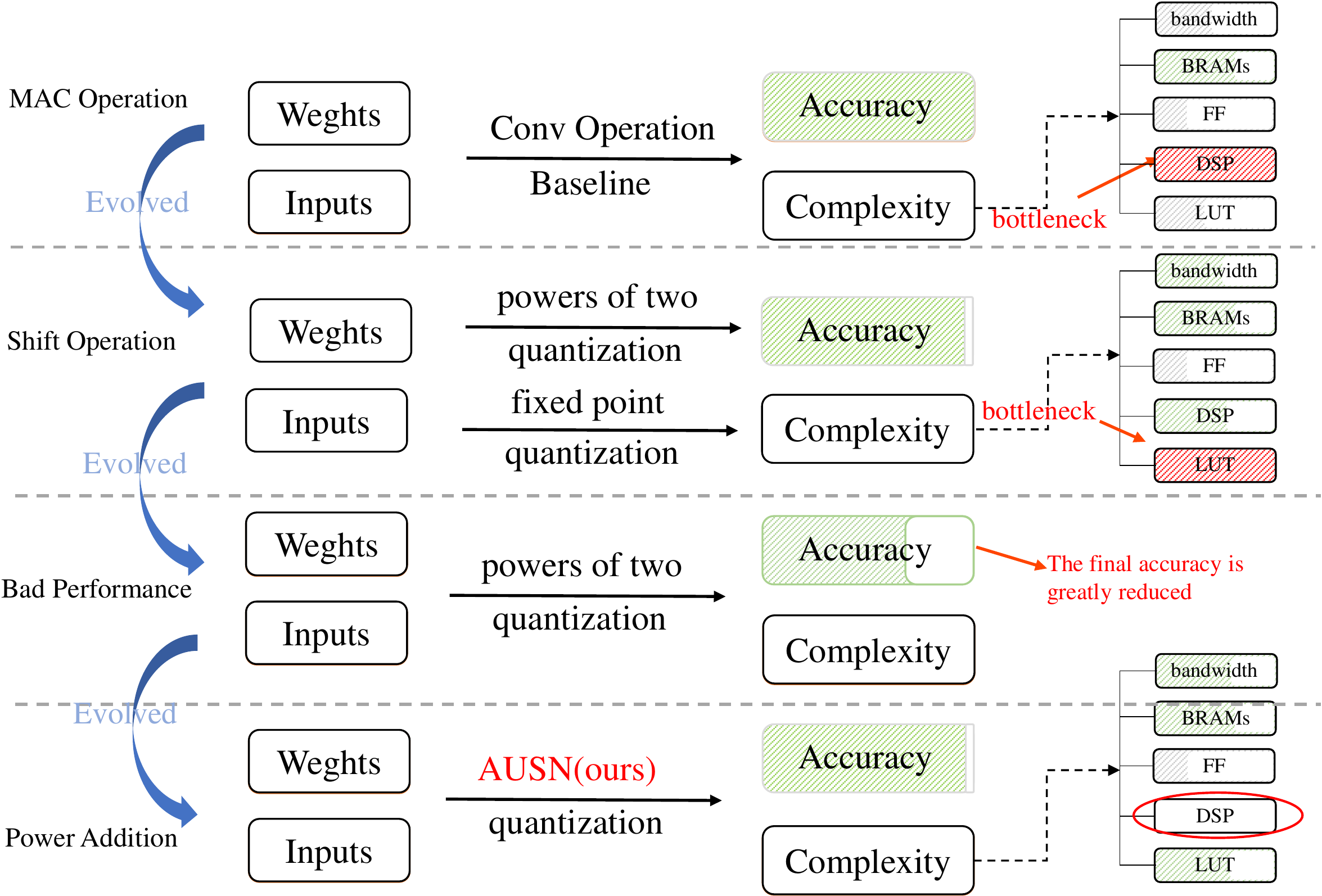}
					\rule[-.5pt]{0pt}{0cm} }
			\end{center}
			\caption{The evaluation of AUSN quantization.}
			\label{fig:evolution}
		\end{figure}

		Most existing quantization methods focus on reducing the bit-width and improve accuracy. However, these approaches ignore the limitation of hardware. As the bit-width of the quantized model decreases, the reductions of the resource and energy consumption manifest themselves differently in various accelerator architectures and arithmetic logics. Fig. 
        \ref{fig:evolution} describes the evaluation method of the proposed AUSN quantization. Generally speaking, multiplication is more expensive than addition and shift operations in terms of the complexity of digital circuits, the number of transistors and the power consumption as well. 
        For example, on FPGA, the shift operation implemented with Look Up Tables (LUTs) is more than $30\times$ more efficient in energy consumption\footnote{$https://china.xilinx.com/support/documentation/sw_manuals/xilinx14_7/ug440-xilinx-power-estimator.pdf$} than the multiplication and accumulation operations conducted with the Digital Signal Processors (DSPs). The DSPs on FPGA are also limited, and the difference between DSP resources and LUT resources is more than two orders of magnitude. 
        
        The power-of-two quantization transforms multiplication operation into shift operation, making the quantized network hardware-friendly to balance the cost of hardware resources. 
        For example, the quantized DNN (the power-of-two quantization quantizes the weights, the input is quantized into the fixed-point number) is deployed on the FPGA, and the multiplication operation can be converted to the shift operation of the fixed-point number.
        Although shift operation can alleviate the shortage of DSP resources to a certain extent, the accumulation operation still depends on DSP and requires additional hardware overhead for the re-quantization. LUT resources may become a new bottleneck because of the bit-width of input. Moreover, due to the significant rounding error caused by the power-of-two quantization, the accuracy will inevitably decrease, so that the power-of-two quantization can not be further applied.
        
        To further reduce the bit-width and computing resources, the weights and activations both are quantized by the power-of-two quantization, which can further convert the shift operation into the addition operation. However, it will lead to a more significant accuracy decrease.
        However, AUSN quantization can not only reduces the accuracy loss but also is reasonable for LUT resources.

		We can quantize weights and activations using proposed AUSN quantization, and transform re-quantization into AUSN rounding. We can eliminate the intermediate value (i.e., partial sums generated in the calculation of convolutional layers) stored on the Block RAMs (BRAMs) for data storage through the logic gate. It can also assure the final accuracy and make full use of LUT resources. In theory, we can even deploy the DNN quantized by AUSN quantization on FPGA without DSP resources or design a reasonable accelerator to realize parallel computing on convolution layer by LUTs and DSPs.

		\subsection{Analysis on LUT Resource}
		The LUT resource consumed by AUSN quantization is far less than that consumed by the shift operation of the fixed-point number as the bit-width of inputs is further reduced. The specific analysis is as follows (take 6bit for both weights and activations as an example).
\begin{figure}[th!] 
			\centering  
				\subfigtopskip=-2pt 
			\subfigure[LUT consumed by the shift operation]{
				\label{fig:LUT-shift}
				\fbox{\rule[-.5pt]{0pt}{0cm} 
					\includegraphics[width=0.5\linewidth]{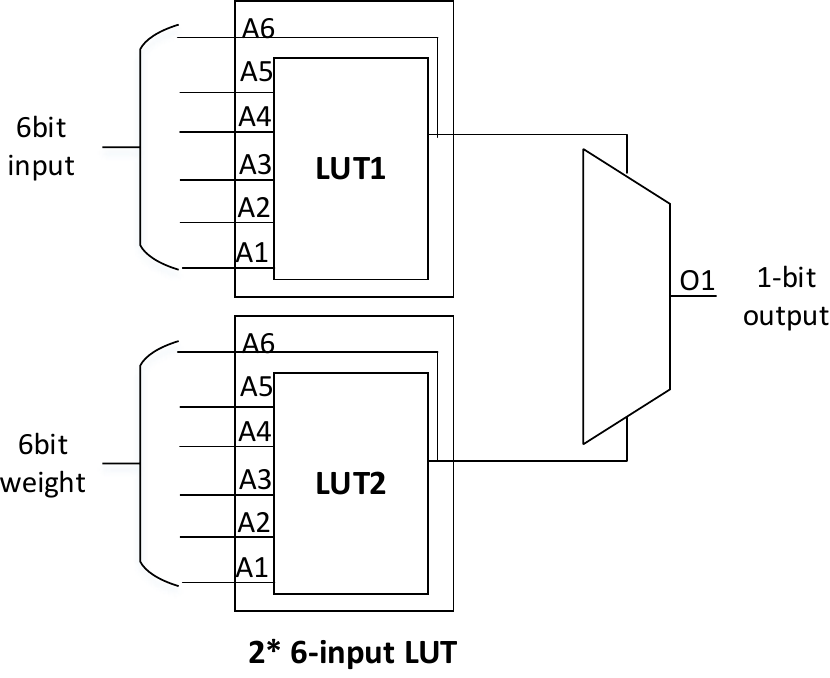}}
				\rule[-.5pt]{0pt}{0cm} }
			
			\subfigure[LUT consumed by the AUSN quantization]{
				\label{fig:LUT-XX}
				\fbox{\rule[-.5pt]{0pt}{0cm} 
					\includegraphics[width=0.75\linewidth]{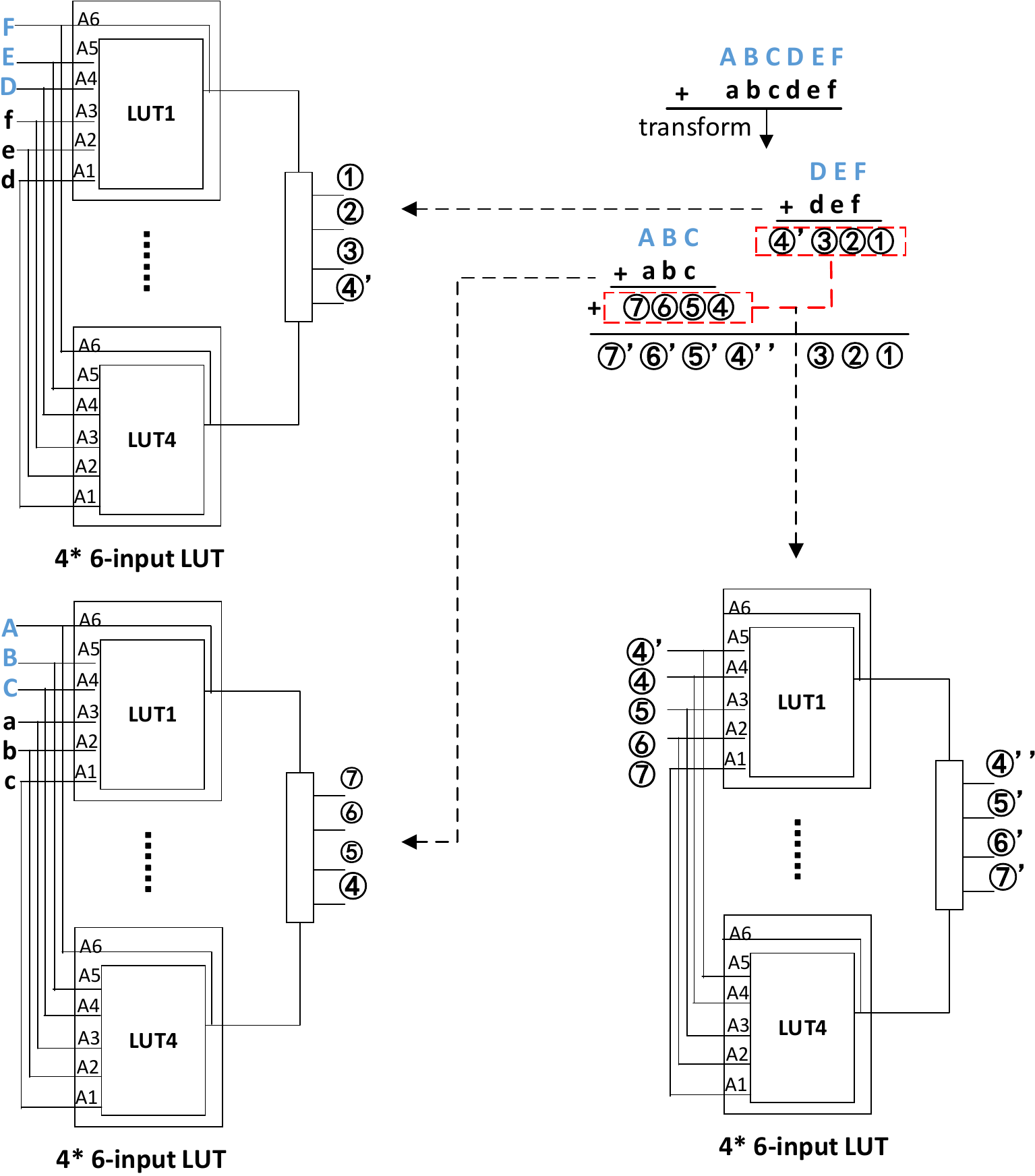}}
				\rule[-.5pt]{0pt}{0cm} }
			\caption{The consumption of LUT resources.}
			\label{fig:consumption-LUT}
		\end{figure}
		
		To make good use of the resources, we implemented addition or shift operation with 6-input LUT (for Modern Xilinx FPGAs, like whole 6 series, 7series). 
        If inputs and weights are quantized to the fixed-point number and power-of-two, respectively, then multiplication of the input and the weight is implemented by the shift operation, the bit-width of the result of the multiplication is 12 bits (6 bits $\times$ 6 bits $\to$ 12 bits). 
        
        In contrast, we quantize the inputs and weights by AUSN quantization, which converts the shift operation of the value into addition operation of the power. We divide the addition operation into 3 times, each addition consuming 4 LUTS. Through this, the result only needs 7 bits (6 bits + 6 bits $\to$ 7 bits). Fig. \ref{fig:consumption-LUT} 
		describes the consumption of LUT resources.
		
		We can found the number of 6-input LUT consumed by the multiplication operation of 6bits implemented by the shift operation on the value (in Fig. \ref{fig:LUT-shift}) is $ 12 \times 2 = 24\,LUTs$, where the number of 6-input LUT consumed by the addition operation on the power (in Fig. \ref{fig:LUT-XX}) is $ 3 \times 4 = 12\,LUTs$. AUSN quantization consumes less LUT resources than the shift operation of fixed-point numbers. Remarkably, there is almost no accuracy loss of the quantized model by AUSN quantization.
		
		\subsection{The analysis on implementation}
			\begin{figure}[hb!] 
    \vspace{-0.2em}
    \centering  
    \setlength{\belowcaptionskip}{-1em}   
    
    \fbox{\rule[-.5pt]{0pt}{0cm} 
    	\includegraphics[width=0.6\linewidth]{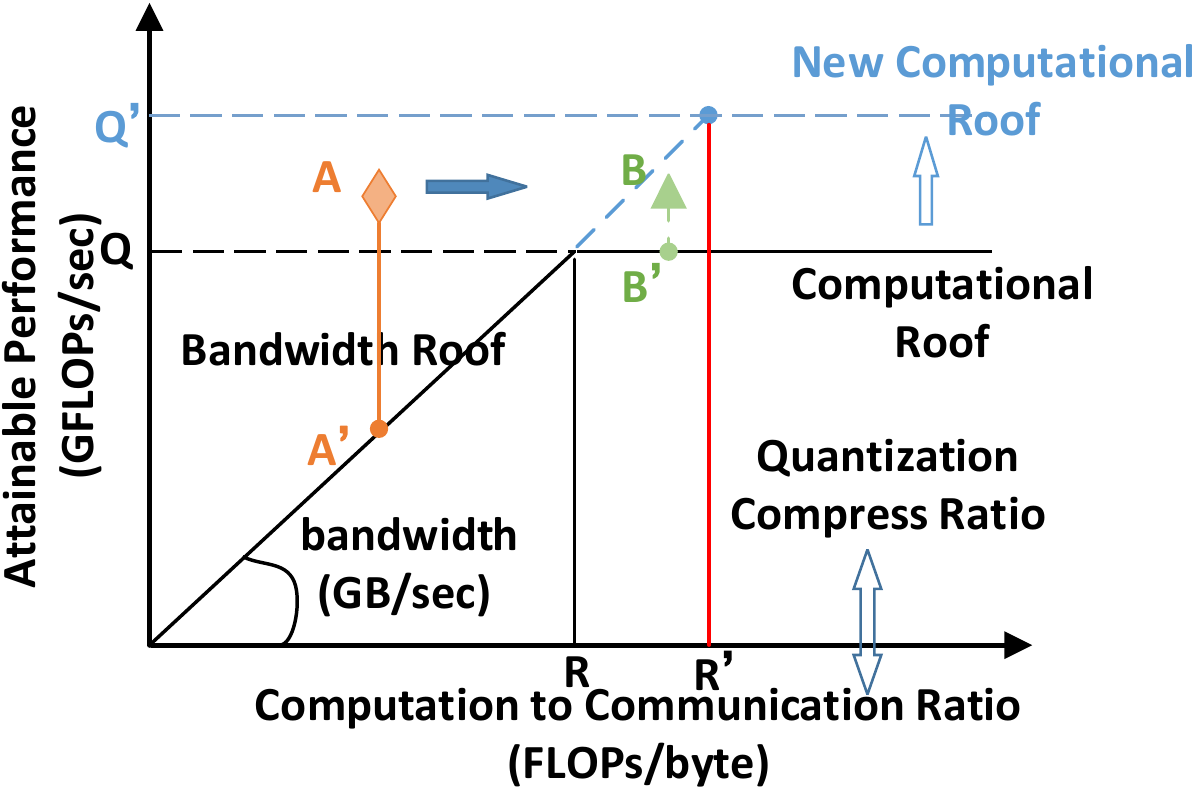}
    	\rule[-.5pt]{0pt}{0cm} }
    \caption{\small Optimization space provided by hardware.}
    \label{fig:rl}
    \end{figure}
	We refer to the classic performance evaluation model, called the RoofLine model~\cite{williams2009roofline}, to analyze the quantization methods and the performance gains of the quantized model on the machine. It can be concluded that quantization reduces the accuracy of operations and operands, resulting in fewer bits representing these operands and corresponding operations, which significantly increase the Computation to Communication Ratio ($CCR$) as defined in Equation (\ref{con:equa6}). 

	\begin{equation}
	\begin{small}
		\begin{aligned}
		\small
		\vspace{-0.8cm}
			\setlength\abovedisplayskip{-1.5em}
			\setlength\belowdisplayskip{-1.5em}
			{Computation\;to\;Communication\;Ratio} &= \frac {FLOPs\;per\;second} {Memory\;access\;per\;second}\\
			&=\frac {total\;number\;of\;operations} {total\;amount\;of\;external\;data\;access}\\
			&=\frac {total\;number\;of\;operations} {4\times\left(weight\;memory + output\;memory\right)}\\
			&=\frac {time\;complexity\;of\;model} {space\;complexity\;of\;model}
			\label{con:equa6}
		\end{aligned}
		\vspace{-0.8em}
	\end{small}
	\end{equation}

	The RoofLine model is the upper bound of theoretical performance that the model and algorithm can achieve under the limitation of the computing power and bandwidth of the accelerator. In the Fig.~\ref{fig:rl}, the x-axis represents the $CCR$ or operational intensity, which the higher $CCR$ is, the higher the memory efficiency is; the y-axis represents the computing performance of the accelerator; $ R$ is the boundary. The bandwidth of machine limits the left area of $R$ and the computing power of the machine limits the right. The upper limit of bandwidth and the upper limit of computation is determined by the accelerator, called bandwidth roof and computational roof. The space/optimization space provided by the machine is the area on the right side of the bandwidth roof. The actual performance can only be the projection on the bandwidth roof or computational roof if it is above the bandwidth roof or computational roof.

	As the model is quantized, the $CCR$ becomes larger, and the position of the model moves to the area to the right of $R$, changing from being limited by bandwidth of accelerator to being limited by computing power of accelerator (transfer from point $A$ to point $B$ in the Fig.~\ref{fig:rl}). Though the bit-width of weights is further reduced, the performance cannot be further improved due to the limitation of computational roof. For FPGA, the computing core is DSPs. The quantized model by the AUSN converts the multiplication operation to shift or even addition operation. These operations can be realized by the LUTs, which is equivalent to adding a computing core to the FPGA, increasing the theoretical calculation peak of FPGA (the computational roof moves up in the Fig.~\ref{fig:rl}), and providing more design space for further optimization.

    Compared with AUSN, the other quantization methods with low bit-width(such as less than 5bit) also can achieve a high compression ratio, it may need to increase the process of \emph{decoding}. The process of \emph{decoding} will increase a lot of hardware resources and energy consumption, which is hardware-unfriendly. As shown in Table~\ref{tab:HLS}, the 4bit shift operation with indexes requires additional \emph{decoders} to be deployed on the hardware, resulting in logic resources and power that even exceed the overhead of 8bit multiplication. Therefore, lowering the bit-width does not indicate the increase of actual benefits.
	\end{appendix}
\end{document}